\documentclass[a4paper,fleqn]{cas-dc}

\usepackage{amsmath,amssymb}   %数学环境
\usepackage[numbers]{natbib}
\usepackage{pdfpages}  %插图
\usepackage{ulem}

\let\printorcid\relax

\begin{document}
		
\title[mode=title]{Primordial Bounce-Inflation Scenario to Alleviate Cosmological Tensions and Lensing Anomaly}
	
\author[1]{Hao-Hao Li}[orcid=0000-0003-0974-771X]
\ead[Hao-Hao Li:]{lihaohao23@hust.edu.cn}

\author[1]{Xin-zhe Zhang}[orcid=0000-0002-3264-7402]
\ead[Xin-zhe Zhang:]{zincz@hust.edu.cn}

\author[1]{Taotao Qiu}
\cormark[1]
\cortext[1]{Corresponding author}
\ead[Taotao Qiu:]{qiutt@hust.edu.cn}

\affiliation[1]{organization={School of Physics, Huazhong University of Science and Technology},
	%	addressline={Jawahar Nagar}, 
	city={Wuhan},
	%               citysep={}, % Uncomment if no comma needed between city and postcode
	postcode={430074}, 
	%	state={Kerala},
	country={China}}

\maketitle

The cosmological tensions, including the Hubble tension and $ S_\text{8} $ tension, have become increasingly serious in the last several years. Moreover, in the Planck-2018 results, it has been found that there is about $10\%$ more lensing than expected in the CMB power spectra, which is called lensing anomaly \cite{Planck:2018vyg}. Instead of the accustomed inflation scenario, this paper presents a Bounce-Inflation (BI) scenario, whose primordial spectrum is more complicated than the standard power-law (PL) form. We will check whether such a scenario can alleviate the Hubble tension, $S_\text{8}$ tension,  and CMB lensing anomaly.

Assuming the standard $ \Lambda $-Cold Dark Matter ($ \Lambda $CDM) cosmological model, cosmic microwave background (CMB) observations show that $ H_\text{0} = 67.4 \pm 0.5\, \text{km}\cdot\text{s}^{-1}\cdot\text{Mpc}^{-1} $ and $ S_\text{8} \equiv \sigma_\text{8} \sqrt{\Omega_\text{m}/0.3} = 0.834 \pm 0.016 $ \cite{Planck:2018vyg}. On the other hand, the local measurements give that $ H_\text{0} = 73.04 \pm 1.04\, \text{km}\cdot\text{s}^{-1}\cdot\text{Mpc}^{-1} $ \cite{Riess:2021jrx} and $S_\text{8} = 0.776 \pm 0.017$ \cite{DES:2021wwk}. There is about $ 5\sigma $ tension of the measurements of $ H_\text{0} $ between CMB under $\Lambda$CDM and local measurements. Similarly, $ 3\sigma $ tension of $ S_\text{8} $ exists between CMB and local measurements. If one uses a parameter $A_\text{L}$ to parameterize the rescaling of the power spectrum due to the lensing effect, the CMB observations give that $A_\text{L} = 1.18 \pm 0.065 $ \cite{Planck:2018vyg}, while in the standard cosmological model, $A_\text{L}=1$. It seems new physics beyond $ \Lambda $CDM is needed because of the irreconcilable cosmological tensions and the emergence of lensing anomaly. There are many extended versions of $ \Lambda $CDM and many of them are differentiated between ``early-time solutions" if the expansion history of our universe is modified before the recombination period and ``late-time solutions" with the modifications of the expansion history after recombination, see \cite{Schoneberg:2021qvd}. 
 
Different from those solutions presented above, here we consider the $\Lambda$CDM universe with a primordial stage described by BI, which we dub as ``primordial time solution".  By such a consideration we're working in the framework of a healthy universe with UV completion, moreover, we don't need to introduce additional modifications to the expansion history of our universe. Hereafter, we will use $ \Lambda$CDM (BI) to represent the $\Lambda$CDM model with a primordial power spectrum of BI and $ \Lambda$CDM (PL) to represent the $\Lambda$CDM model with a PL primordial power spectrum. Usually, the $ \Lambda$CDM (PL) is called the standard cosmological model. To investigate whether the BI scenario can remit CMB lensing anomaly, we also take into account an extended model with $ \Lambda \text{CDM (BI)} + A_\text{L} $.

 BI scenario is an interesting alternative to the standard Big Bang with inflation scenario. By having a non-singular bounce before inflation, such a scenario can successfully avoid the initial singularity that plagued the early evolution history of our universe. Moreover, a lot of progress in the literature has shown that such a scenario can also solve the Big Bang problems such as flatness, horizon, relics, and entropy\cite{Ijjas:2021zwv}. Although many concrete models are realizing BI scenarios as we show one of them in Supplementary Material \href{Sup_Mat}{A}, in the text we parameterize the background of the BI scenario and calculate the primordial power spectrum based on the effective field theory \cite{Ni:2017jxw}.  The parameterized form of the scale factor is
\begin{equation}
	a  = \begin{cases}
		a_{\text{con}}\left(\tilde{\eta}_{\text{B-}}-\eta\right)^{\frac{1}{\epsilon_{\text{c}}-1}} &  \eta<\eta_{\text{B-}},\\
		a_{\text{B}}\left[1+\frac{\alpha}{2}\left(\eta-\eta_{\text{B}}\right)^{2}\right] & \eta_{\text{B-}} \leq \eta \leq \eta_{\text{B+}}, \\ 
		a_{\text{exp}}\left(\tilde{\eta}_{\text{B+}}-\eta\right)^{\frac{1}{\epsilon_{\text{e}}-1}} & \eta>\eta_{\text{B+}}.
	\end{cases}
 \label{scalefactor}
\end{equation}
 where $\eta$ is the conformal time. This parametrization gives rise to the curvature power spectrum as:
\begin{equation}
	\begin{aligned}
		P_{ \mathcal{R}} = \frac{ H_{\text{exp}} ^2}{ 8 \pi ^2 M_\text{p} ^2 \epsilon_\text{e}}\left(\frac{k}{k_\text{*}} \right)^{3-2\nu_\text{e}} \left| C_\text{1} - C_\text{2}\right| ^2 \label{PS-BI}
	\end{aligned},
\end{equation}
where in addition to the standard power spectrum of inflation, the factor $\left| C_\text{1} - C_\text{2}\right| ^2$ is added due to the effect of the preceding bouncing process (see the detailed calculations and resultant form in Supplementary Material \href{Sup_Mat}{A}). 
Then we use the package \textsf{CLASS}\footnote{\url{https://lesgourg.github.io/class_public/class.html}} as the Einstein-Boltzmann equation solver and the package \textsf{MontePython}\footnote{\url{https://baudren.github.io/montepython.html}} as the Markov Chain Monte Carlo(MCMC) sampler to product the post prior of cosmological parameters. The data sets we used are \texttt{Planck 2018: Planck\_highl\_TTTEEE, Planck\_lowl\_EE, Planck\_lowl\_TT, Planck\_lensing} (hereafter, \texttt{P18}) \cite{Planck:2018vyg}
and \texttt{SPT-3G 2018: SPT-3G TTTEEE} (hereafter, \texttt{SPT3G}) \cite{SPT-3G:2022hvq}.  To plot the posterior distributions, we use \textsf{GetDist}\footnote{\url{https://getdist.readthedocs.io}}. The Gelman-Rubin criterion for all chains is converged to $ R-1 < 0.01 $. Our numerical materials have been uploaded to GitHub \footnote{\url{https://github.com/Haohaolic/bounce_inflation}}.

In our numerical results, based on flat $\Lambda$CDM, the BI scenario gives the results that $ H_\text{0} = 68.60^{+0.40}_{-0.45}\, \text{km}\cdot\text{s}^{-1}\cdot\text{Mpc}^{-1}$ by using \texttt{P18} data sets and $ H_\text{0} = 68.96 \pm 0.38 \, \text{km}\cdot\text{s}^{-1}\cdot\text{Mpc}^{-1}$ by using \texttt{P18} + \texttt{SPT3G} data sets.  These reduce the Hubble tension slightly, but it is still serious ($ > 3 \sigma $ level).  In the extended $ \Lambda \text{CDM (BI)} + A_\text{L} $ model, the results are $ H_\text{0} = 69.38 \pm 0.49 \, \text{km}\cdot\text{s}^{-1}\cdot\text{Mpc}^{-1}$ fitted by \texttt{P18} data sets and $ H_\text{0} = 69.49 \pm 0.45 \, \text{km}\cdot\text{s}^{-1}\cdot\text{Mpc}^{-1}$ fitted by \texttt{P18} + \texttt{SPT3G} data sets (see Figure \ref{Fig.1}, also see the Table II of the
Supplementary Material \href{Sup_Mat}{B} ), which reduce the Hubble tension to $\sim 3\sigma $ level compared with the local observational result $ H_\text{0} = 73.04 \pm 1.04\, \text{km}\cdot\text{s}^{-1}\cdot\text{Mpc}^{-1}  $ \cite{Riess:2021jrx}. 

\begin{figure}[htbp] %H为当前位置，!htb为忽略美学标准，htbp为浮动图形
	\centering %图片居中
	\includegraphics[width=0.45\textwidth]{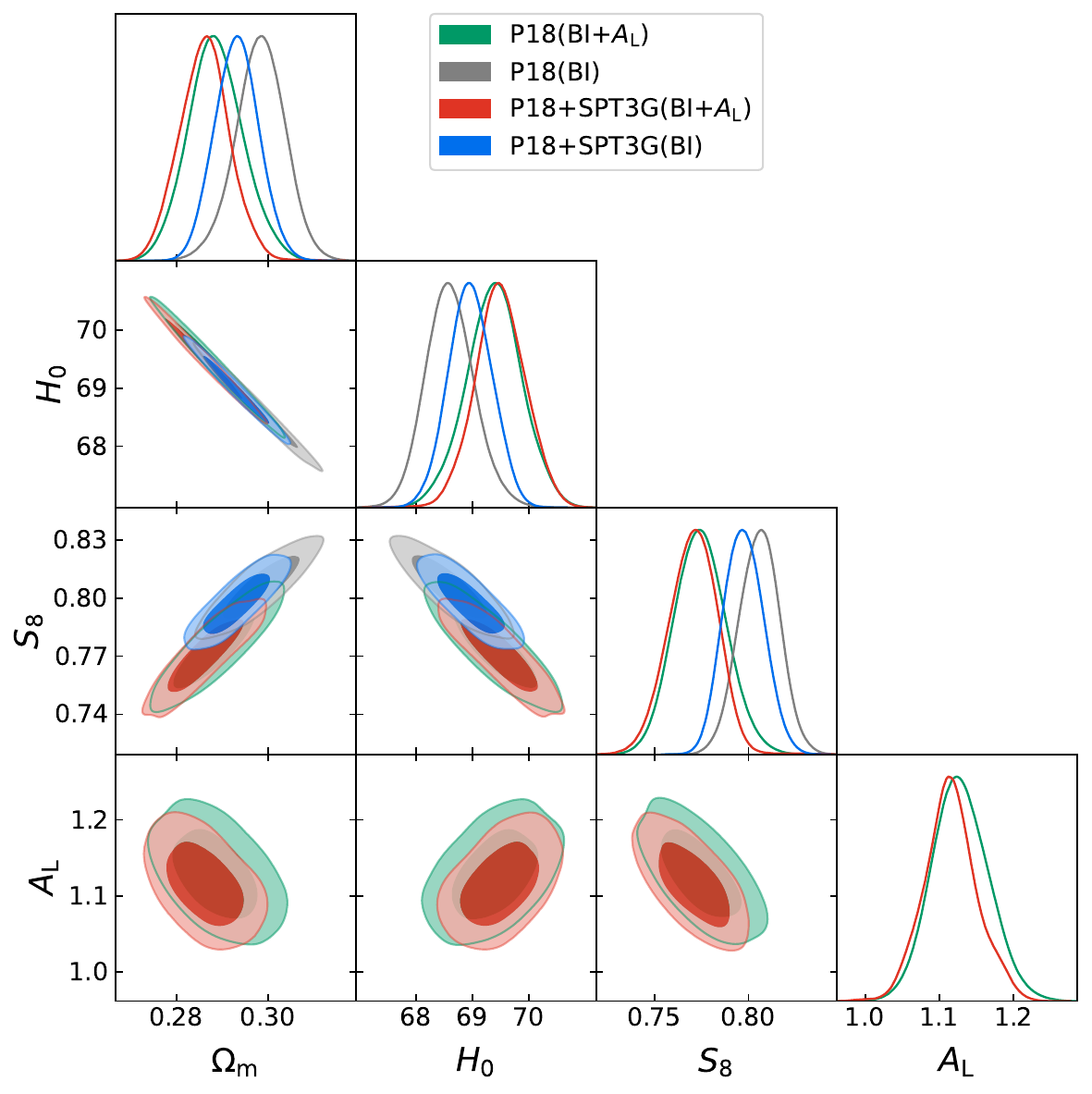} %插入图片，[]中设置图片大小，{}中是图片文件名
	\caption{The posterior distributions of $H_\text{0}$, $S_\text{8} $, $ \Omega_\text{m} $ and $ A_\text{L} $. Notice that in the $ \Lambda\text{CDM (BI)}$ model, $A_\text{L} =1$.} %最终文档中希望显示的图片标题
	\label{Fig.1} %用于文内引用的标签
\end{figure}%结束环境

Similarly, in the $ \Lambda\text{CDM (BI)} $ model, our results give that $ S_\text{8} = 0.806 \pm 0.011 $ by using \texttt{P18} data sets and $ S_\text{8} = 0.797\pm 0.010 $ by using \texttt{P18} + \texttt{SPT3G} data sets. There is still $ 2\sigma $ tension of $S_\text{8}$  between the values obtained from CMB and the values measured directly from local observations \cite{DES:2021wwk}. When we consider the $ \Lambda \text{CDM (BI)} + A_\text{L} $ model, the results are $ S_\text{8} = 0.774 \pm 0.014 $ fitted by \texttt{P18} data sets and $ S_\text{8} = 0.771^{+0.013}_{-0.012} $ fitted by \texttt{P18} + \texttt{SPT3G}, which show {\it no $S_\text{8} $ tension} at $ 1 \sigma $ level compared with the local observational result  $S_\text{8} = 0.776 \pm 0.017$  \cite{DES:2021wwk}.  Since $ S_\text{8} $ is a combination parameter to describe the growth of matter fluctuations, a smaller $ S_\text{8} $ expresses a lower value of the matter density parameter (see Figure \ref{Fig.1}, also see Table II of the Supplementary Material \href{Sup_Mat}{B}). The $ \Omega_\text{m}  = 0.2983 \pm 0.0055$ in our result based on $ \Lambda $CDM (BI) is smaller than the matter density parameter of the Planck-2018 result that $\Omega_\text{m} = 0.315 \pm 0.007$ based on $ \Lambda $CDM (PL) \cite{Planck:2018vyg}. 
 
Another important results are that the spectral indices $ n_\text{s} \approx 0.980 $ (\texttt{P18}) and $ n_\text{s} \approx 0.984 $ (\texttt{P18} + \texttt{SPT3G}) in the both $ \Lambda\text{CDM (BI)} $ and $ \Lambda \text{CDM (BI)} + A_\text{L} $ models. These are larger than the spectral index of the Planck-2018 result which $n_\text{s} \approx 0.965  $ \cite{Planck:2018vyg} (see the Table II of the Supplementary Material \href{Sup_Mat}{B} ). Our results show a trend in the scale-invariance spectrum named the Harrison-Zel'dovich spectrum. This gets the same conclusions as the early dark energy scenario to solve the Hubble tension \cite{Ye:2022efx}. Planck-2018 result shows an $ 8\sigma $ tilt away from scale-invariance for the base-$ \Lambda $CDM (PL), but it seems that a larger spectral index is helpful to solve cosmological tensions.

Moreover, we also investigate the possibility of releasing weak lensing in the BI scenario. Weak lensing of the CMB smooths the shape of the observed power spectra. The gravitational lensing amplitude is defined as a rescaling factor of lensing potential power-spectrum \cite{Calabrese:2008rt}:
$	C_{\ell}^{\phi\phi} \longrightarrow A_\text{L} C_{\ell}^{\phi\phi} $, 
where $ A_\text{L} = 0 $ implies that the CMB is unlensed while $ A_\text{L} = 1 $ is the expected value in $ \Lambda $CDM (PL) model. Our results show that $A_\text{L}$ can be reduced to $ 1.128 \pm 0.038 $ with \texttt{P18} data sets and $ 1.116 \pm 0.036 $ with \texttt{P18} + \texttt{SPT3G} data sets in the $ \Lambda\text{CDM (BI)}+A_\text{L} $ model (see the Table II of the
Supplementary Material \href{Sup_Mat}{B} ). Since the $\Lambda$CDM (BI) model enlarges the parameter space concerning the $\Lambda$CDM (PL) model, degeneracy between the power spectrum template and the $A_\text{L}$ parameter may arise. 
 
\begin{figure}[htbp] %H为当前位置，!htb为忽略美学标准，htbp为浮动图形
	\centering %图片居中
	\includegraphics[width=0.45\textwidth]{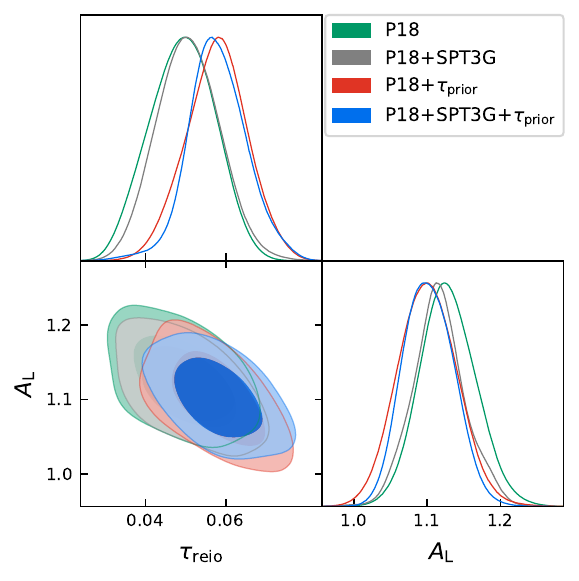} %插入图片，[]中设置图片大小，{}中是图片文件名
	\caption{In $ \Lambda\text{CDM (BI)} + A_\text{L} $ model, optical depth $ \tau_{\text{reio}} $ and the gravitational lensing amplitude $ A_\text{L} $ with and without the likelihood $\tau_{\text{prior}} = 0.088 \pm 0.015 $.} %最终文档中希望显示的图片标题
	\label{Al-tau} %用于文内引用的标签
\end{figure}

It is known that the reionization of the universe is another effect that washes out the primordial anisotropies and the  CMB power spectrum is suppressed by a factor of $ e^{-2\tau_{\text{reio}}}$. Thus there is a strong degeneracy between optical depth $ \tau_{\text{reio}} $ and gravitational lensing amplitude $ A_\text{L} $. We guess the CMB lensing anomaly may be caused by the degeneracy between $ \tau_{\text{reio}} $ and $ A_\text{L} $. So we add a Gaussian prior of the optical depth with $ \tau_{\text{prior}} = 0.088 \pm 0.015 $ from WAMP 9-year results \cite{WMAP:2012nax} as a new likelihood artificially. Our result implies that data sets with a prior of optical depth will give a larger $ \tau_{\text{reio}} $ that is approximately equal to 0.058 and the smaller $ A_\text{L} $ that is approximately equal to 1.10, which will ulteriorly remit the CMB lensing anomaly, as shown in Figure \ref{Al-tau}.  By the way, the cosmological tensions are not aggravated when we add this prior ( see Table III of the
Supplementary Material \href{Sup_Mat}{C} ).

%结束环境

The $\chi^2 $ goodness-of-fit tests for BI with different data sets are shown in Table \ref{Table-1}. The reference model is our best-fit $\Lambda\text{CDM}$ (PL) cosmology from the \texttt{P18} or \texttt{P18}+\texttt{SPT3G} likelihood combination. The $ \Delta \chi^2 $ is defined as $ \Delta \chi^2  = \chi^2 - \chi^2_{\Lambda\text{CDM(PL)}, \text{P18}} $ if only \texttt{P18} data sets used or $ \Delta \chi^2  = \chi^2 - \chi^2_{\Lambda\text{CDM(PL)}, \text{P18} +\text{SPT3G}} $ if \texttt{P18}+\texttt{SPT3G} data sets combined. In all cases, the $ \Delta \chi^2 $ is greater than 0 but not by too much, so the general agreement of the data with the cosmological model BI is reasonable.

\begin{table}[htbp]
	\centering
	\caption{The goodness-of-fit tests for BI or BI+$A_\text{L}$ with different data sets. The fiducial values we have run are $ \chi^2_{\Lambda\text{CDM(PL)}, \text{P18}} = 2780.72 $, $ \chi^2_{\Lambda\text{CDM(PL)}, \text{P18}+\text{SPT3G}} = 4671.02 $.}
	\begin{tabular} { |l | c| c |}
		\hline
		Data sets (Model) & $\chi^2$  &  $\Delta\chi^2 $  \\
		\hline
		P18 (BI) & $2794.14           $ & $13.42            $ \\
		
		P18 (BI+$A_\text{L}$) & $2783.98     $ & $3.26        $\\
		
		P18+SPT3G (BI) & $4705.76        $ & $34.74        $\\
		
		P18+SPT3G (BI+$A_\text{L}$) & $4692.44        $ & $21.42    $ \\
		\hline 
	\end{tabular}
	\label{Table-1}
\end{table}

In summary, this work shows that the primordial scenario can alleviate cosmological tensions and CMB lensing anomaly. By utilizing the BI scenario, several achievements are obtained: 
(1) The Hubble tension is reduced based on the $ \Lambda$CDM (BI) model. If we consider the extension of the $ \Lambda $CDM (BI) model with gravitational lensing amplitude $ A_\text{L} $, there is a nearly $ 3\sigma $ tension (compared to $>5\sigma$ before) of the Hubble parameter compared to local measurements. (2) The $ S_\text{8} $ tension is reduced to nearly $ 2\sigma $ level in the $ \Lambda $CDM (BI) model. Moreover, $ S_\text{8} $ with measurements of CMB in the $ \Lambda\text{CDM (BI)} + A_\text{L} $ model are consistent with the local measurements at $ 1\sigma $ level. We also show that in this case, it tends to reach a Harrison-Zel'dovich primordial power spectrum. (3) The CMB lensing anomaly is remitted compared with the result of Planck 2018. It is also found that breaking the degeneracy between optical depth $ \tau_{\text{reio}} $ and lensing amplitude $ A_\text{L} $ is helpful to alleviate the CMB lensing anomaly. 

 As a final remark, although in this work we're focusing only on Planck+SPT3G data, it is interesting to extend our study with more datasets such as ACT, DESI or JWST. Due to the page limit, we postpone such kind of discussions for an upcoming work.

\section*{Conflict of interest}
The authors declare that they have no conflict of interest.

\section*{Acknowledgments}
We thank Yun-Song Piao, Jun-Qian Jiang and Jun-Qing Xia for their helpful discussions. T.Q. acknowledges the Institute of Theoretical Physics, Chinese Academy of Sciences for its hospitality during his visit there. T.Q. is supported by the National Key Research and Development Program of China (Grant No. 2021YFC2203100), as well as Project 12047503 supported by the National Natural Science Foundation of China. H.-H.L. wishes to acknowledge the support of the China Postdoctoral Science Foundation, GZC20230902. 

\section*{Author contributions}
Taotao Qiu conceived the idea. He also initiated this study with all other authors. Hao-Hao Li and Xin-zhe Zhang conducted numerical calculations. Hao-Hao Li analyzed the physical results and wrote the manuscript.  All authors discussed the results together.

\appendix
\section*{Appendix. Supplementary materials}

Supplementary materials to this short communication can be
found online at \hyperlink{https://doi.org/10.1016/j.scib.2025.01.007}{https://doi.org/10.1016/j.scib.2025.01.007}.

%\bibliographystyle{apsrev4-2}
%\bibliography{ref.bib}% Produces the bibliography via bibTeX.

\includepdf[pages={1-7}]{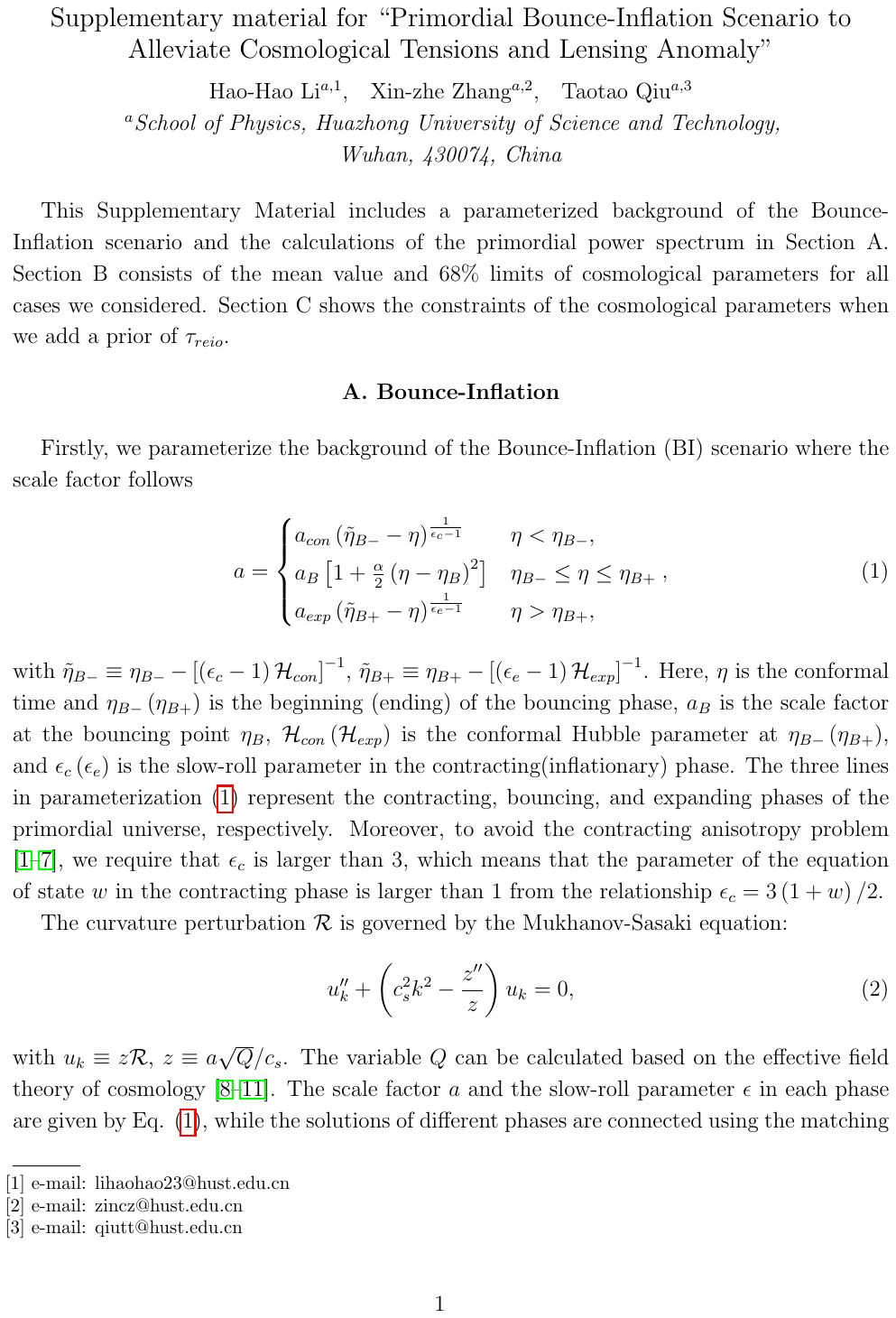}

\end{document}